\documentstyle[prd,tighten,aps]{revtex}
\begin{document}
\twocolumn[\hsize\textwidth\columnwidth\hsize
\csname@twocolumnfalse\endcsname
\title{Quantization of electric charge, the neutrino and generation 
       nonuniversality}
\author{A. Doff and F. Pisano}
\address{Departamento de F\'\i sica, Universidade Federal do Paran\'a 
81531-990 Curitiba PR, Brazil}
\date{\today}
\maketitle
\begin{abstract}
It is showed that the electric charge quantization is unconnected to 
Majorana neutrino in the non-universal generations leptoquark-bilepton 
flavordynamics which includes the right-handed neutrino and an explicit 
U(1) factor in the gauge semisimple group.
\end{abstract}
\pacs{PACS numbers: xxxx}
\vskip2pc] 
\newpage 
Two fundamental questions leaved open within the standard model of 
nuclear and electromagnetic interactions~\cite{GWS} are the flavor 
question and the fermion mass hierarchy, being 
addressed~\cite{PP,Frampton92,fp96,DP00,Tullyetal} in the 
leptoquark--bilepton flavordynamics~\cite{PP,Frampton92,Footetal} having 
the local gauge symmetries 
$
G_{3m1}\equiv {\rm SU(3)}_c\otimes {\rm SU}(m)_L\otimes {\rm U(1)}_{L+R}
$
with $m=3,4$. 
There is also a cubic seesaw relation $m(\nu_\ell) = m^3(\ell)/M_W^2$ 
constraining the neutrino mass to the charged lepton masses~\cite{framea} 
implicating the $10^{-5}$ eV value for the lightest neutrino mass, 
and also the interesting possibility of double beta decay depending less 
on the neutrino mass~\cite{PlTon} than in many extensions of the 
standard model, as well as the associated Majoron emission 
proccess~\cite{euShelly}. 
It was also showed the non-trivial issue that in the 
$m=3$ model there is a Peccei--Quinn symmetry~\cite{PQ77} with an 
invisible axion, solving the strong-CP problem~\cite{Pal}. 
Although in the minimal $G_{331}$ leptoquark-bilepton 
model the massive neutrinos are Majorana fermions~\cite{OLR,OkYa}, 
due to the non-universal generations 
structure the electric charge quantization and Majorana neutrino connection 
is lost. Such fact agrees with the pioneering scrutiny~\cite{BM} involving 
a large class of gauge models containing a U(1) factor in the gauge 
group, the right-handed neutrino and generations universality. 
Using the lightest leptons as the particles which determine the 
approximate symmetry with the generations nonuniversality and if 
lepton charges are only $0$, $\pm 1$, SU(4) is the highest symmmetry 
group to be included in the electroweak sector. There is no room for 
$m>4$. A model with the SU(4)$\otimes$U(1) symmetry was proposed more 
than one decade ago by Voloshin~\cite{Voloshin} but quarks were not 
included there. 
\par
Let us consider the largest leptoquark--bilepton gauge semisimple 
$G_{341}$ group extension. 
The electric charge operatorial formula~\cite{pp95}
\begin{equation}
\frac{{\cal Q}}{|e|} = 
\Lambda_3 + \xi\Lambda_8 + \zeta\Lambda_{15} + \varsigma N\Lambda_0
\label{uno}
\end{equation}
is embedded in the traceless neutral generators $\Lambda_i$, 
$i=3,8,15$ of the SU(4) gauge group,
\begin{mathletters}
\begin{eqnarray}
\Lambda_3 & = & \frac{1}{2}{\rm diag}(1,-1,0,0), \\
\label{dueuno}
\Lambda_8 & = & \frac{1}{2\sqrt 3}{\rm diag}(1,1,-2,0), \\
\label{dudue}
\Lambda_{15} & = & \frac{1}{2\sqrt 6}{\rm diag}(1,1,1,-3) 
\label{dutre}
\end{eqnarray}
\end{mathletters}
and 
$\Lambda_0 = {\rm diag}\,(1,1,1,1)\,,$ 
with the specific embedding parameters $\xi=-1/{\sqrt 3}$, 
$\zeta = -2{\sqrt 6}/3$, $\varsigma =1$, where $N$ is the new charge 
associated to the symmetric non-chiral Abelian semisimple factor of 
the $G_{341}$ local symmetry. 
In the $\bar{\bf 4}$ representation the neutral 
generators are~\cite{MPP94} 
\begin{mathletters}
\begin{eqnarray}
\bar\Lambda_3 & = & \frac{1}{2}{\rm diag}(0,0,1,-1), \\ 
\label{treuno}
\bar\Lambda_8 & = & \frac{1}{2\sqrt 3}{\rm diag}(0,2,-1,-1), \\
\label{tredue}
\bar\Lambda_{15} & = & \frac{1}{2\sqrt 6}{\rm diag}(3,-1,-1,-1).
\label{tretre}
\end{eqnarray}
\end{mathletters}
The SU(4) maximal subalgebras are ${\rm SU}(3)\otimes {\rm U}(1)$, 
${\rm SU}(2)\otimes{\rm SU}(2)\otimes {\rm U}(1)$, ${\rm Sp}(2)$, 
and ${\rm SU}(2)\otimes{\rm SU}(2)$.  
The isomorphism is ${\rm SU}(4)\sim{\rm SO}(6)$. 
Three families of independent leptonic chiral flavor gauge symmetry 
eigenstates tranforms collectively, 
\begin{equation}
f_{\ell L} = \left (
\begin{array}{c}
\left (
\begin{array}{c}
\nu_{\ell L} \\ 
\ell_L 
\end{array}
\right ) \\
\left (
\begin{array}{c}
\nu_{\ell R} \\ 
\ell_R
\end{array}
\right )^c
\end{array}
\right ) \sim ({\bf 1}_c,{\bf 4}_L, N = 0)
\label{um}
\end{equation}
with the label $\ell = e, \mu, \tau$ and the charge conjugated fields 
$\ell^c= C \bar\ell^{\rm T}$, $\nu^c_\ell$, being 
$C\equiv i\gamma^2\gamma^0$. One quark family has the attributions 
\begin{equation}
Q_{1L} = \left (
\begin{array}{c}
u_1 \\ 
d_1 \\ 
u^\prime \\ 
J
\end{array}
\right )_L \sim ({\bf 3}_c,{\bf 4}_L,+2/3 )
\label{dois}
\end{equation}
with the singlets 
\begin{mathletters}
\begin{eqnarray}
u_{1R} & \sim & ({\bf 3}_c,{\bf 1}_R,+2/3), \\
d_{1R} & \sim & ({\bf 3}_c,{\bf 1}_R,-1/3), \\
u^\prime_R & \sim & ({\bf 3}_c,{\bf 1}_R,+2/3), \\
J_R & \sim & ({\bf 3}_c, {\bf 1}_R, +5/3),
\end{eqnarray}
\end{mathletters}
and the other two families have the following common transformation 
properties 
\begin{equation}
Q_{\alpha L} = \left (
\begin{array}{c}
j_\alpha \\ 
d^\prime_\alpha \\ 
u_\alpha \\ 
d_\alpha
\end{array}
\right )_L \sim ({\bf 3}_c,\bar{{\bf 4}}_L,-1/3), 
\label{tres}
\end{equation}
$\alpha = 2,3$ for the chiral left-handed fields and 
\begin{mathletters}
\begin{eqnarray}
j_{\alpha R} & \sim & ({\bf 3}_c,{\bf 1}_R,-4/3), \\
\label{quaun}
d^\prime_{\alpha R} & \sim & ({\bf 3}_c,{\bf 1}_R,-1/3), \\ 
\label{qudu}
u_{\alpha R} & \sim & ({\bf 3}_c,{\bf 1}_R,+2/3), \\ 
\label{qutre}
d_{\alpha R} & \sim & ({\bf  3}_c,{\bf 1}_R,-1/3),
\label{ququa}
\end{eqnarray}
\end{mathletters}
where $u^\prime$, $J$, $j_{\alpha}$ and $d^\prime_\alpha$ are new 
quark flavors with electric charges $+\frac{2}{3}|e|$, $+\frac{5}{3}|e|$, 
$-\frac{4}{3}|e|$ and $-\frac{1}{3}|e|$ respectively, where the down-like 
quark flavors transport the quantum of electric charge being
\begin{eqnarray}
|e| & = & \frac{g\,t}{(1 + 4t^2)^{1/2}} = \frac{g^\prime}{(1 + 4t^2)^{1/2}} 
          \nonumber \\
    & = & 1.602\,176\,462(63)\times 10^{-19}\,\,{\rm C} \nonumber \\
    & = & 4.803\,204\,20(19)\times 10^{-10}\,\,{\rm esu}
\label{qquuaa}
\end{eqnarray}
the proton charge~\cite{PDG2000}, and $t\equiv g^\prime/g$, where $g$ and 
$g^\prime$ are the SU(4) and U(1) gauge coupling constants. 
The electric charged and neutral leptons acquire mass through the 
symmetric decuplet $({\bf 1}_c,\overline{{\bf 10}}_S, 0)$ of scalar 
fields,
\begin{equation}
\bar{H} = \left (
\begin{array}{cccc}
H^0_1 & H^+_1 & H^0_2 & H^-_2 \\ 
H^+_1 & H^{++}_1 & H^+_3 & H^0_3 \\ 
H^0_2 & H^+_3 & H^0_4 & H^-_4 \\ 
H^-_2 & H^0_3 & H^-_4 & H^{--}_2
\end{array}
\right )
\label{cinq}
\end{equation}
in the Yukawa interactions 
\begin{equation}
{\cal L}^f_{\rm Y} = -\frac{1}{2}G_{\ell\ell^\prime}
\overline{(f_{\ell L})^c} \bar{H} f_{\ell^\prime L}
\label{seii}
\end{equation}
having the general form of a Majorana mass term after the spontaneous 
symmetry breaking. Since the lepton mass term transforms as 
$\overline{(f_{\ell L})^c} f_{\ell^\prime R} 
\sim ({\bf 1}_c,{\bf 4}\otimes {\bf 4},0) = 
({\bf 1}_c,{\bf 6}_A\oplus {\bf 10}_S,0)$ 
and the sextet will leave one lepton massless and two others degenerate 
for three families, it is necessary to introduce the 
$N=0$ symmetric decuplet which plays no role in generating quark masses. 
The explicit fermion bilinears and Higgs 
bosons Yukawa couplings are 
\begin{eqnarray}
&{}&\overline{(f_{L})^c}\bar{H} f_L  =  \nonumber \\
&\mbox{}&\overline{(\nu^c_\ell)}_R \nu_{\ell R} H_1^0 + 
\overline{(\ell^c)}_R\nu_{\ell L} H_1^{+} + 
\bar\nu_{\ell R}\nu_{\ell L} H_2^0 + 
\bar\ell_R\nu_{\ell L} H_2^- \nonumber \\
& + & 
\overline{(\nu^c_\ell)}_R\ell_L H_1^+ + 
\overline{(\ell^c)}_R\ell_L H_1^{++} + 
\bar\nu_{\ell R}\ell_L H_3^+ + 
\bar\ell_R\ell_L H^0_3 \nonumber \\
& + & 
\overline{(\nu^c_\ell)}_R\nu^c_{\ell L} H_2^0 + 
\overline{(\ell^c)}_R\nu^c_{\ell L} H^+_3 + 
\bar\nu_{\ell R}\nu^c_{\ell L} H^0_4 + 
\bar\ell_R\nu^c_{\ell L} H^-_4 \nonumber \\ 
& + & 
\overline{(\nu^c_\ell)}_R\ell^c_L H^-_2 + 
\overline{(\ell^c)}_R\ell^c_L H^0_3 + 
\bar\nu_{\ell R}\ell^c_L H^-_4 + 
\bar\ell_R\ell^c_L H^{--}_2 \nonumber \\
\label{sette}
\end{eqnarray}
and after the spontaneous symmetry breaking steps
\begin{equation}
G_{341}\rightarrow G_{331}\rightarrow G_{321}\rightarrow G_{31}
\label{otto}
\end{equation}
remain the following mass terms, 
\begin{eqnarray}
-{\cal L}_{\rm Y}^{\rm mass} & = & \frac{1}{2}
(G_{\nu_\ell\nu_{{\ell^\prime}}}\langle H^0_1\rangle_0
\overline{(\nu^c_\ell)}_R\nu_{\ell^\prime L} + 
G_{\nu_{\ell}\nu_{\ell^\prime}}\langle H^0_2\rangle_0
\bar\nu_{\ell R}\nu_{\ell^\prime L} \nonumber \\
& + & 
G_{\ell\ell^\prime}\langle H^0_3\rangle_0\bar\ell_R\ell^\prime_L + 
G_{\nu^c_\ell\nu^c_{\ell^\prime}}\langle H^0_2\rangle_0
\overline{(\nu^c_\ell)}_R\nu^c_{\ell^\prime L} \nonumber \\
& + & 
G_{\nu_\ell\nu_{\ell^\prime}}\langle H^0_1\rangle_0
\bar\nu_{\ell R}\nu^c_{\ell^\prime L} + 
G_{\ell\ell^\prime}\langle H^0_3\rangle_0\overline{(\ell^c)}_R\ell^c_L)
\end{eqnarray}
and the neutrinos are Dirac--Majorana particles~\cite{Espetal}. 
\par
The vanishing anomalies conditions containing the U(1)$_N$ fermionic 
attributions imply the following constraints between the $N$'s:
\begin{mathletters}
\begin{eqnarray}
{\rm Tr}([{\rm SU}(3)_c]^2[{\rm U}(1)_N]) & = & 0: \nonumber \\
3(N_{Q_{1L}} + 2N_{Q_{\alpha L}}) - 
3(N_{U_{1,\alpha R}} + N_{D_{1,\alpha R}}) \nonumber \\ 
- N_{J_R} - 2N_{j_{\alpha R}} - N_{u^\prime_R} 
- 2N_{d^\prime_{\alpha R}}) & = & 0, \\
\label{uunno}
{\rm Tr}([{\rm SU}(4)_L]^2[{\rm U}(1)_N]) & = & 0 : \nonumber \\ 
3(N_{Q_{1 L}} + 2 N_{Q_{\alpha L}}) + \sum_\ell N^3_\ell & = & 0, \\
\label{dduee}
{\rm Tr}([{\rm U}(1)_N]^3) & = & 0: \nonumber \\
3(N^3_{Q_{1L}} + 2N^3_{Q_{\alpha L}}) - 
3(N^3_{U_{1,\alpha R}} + N^3_{D_{1,\alpha}}) + \nonumber \\
N^3_{J_R} - 2N^3_{j_{\alpha R}} - N^3_{u^\prime_R}
-2N^3_{d^\prime_{\alpha R}} + \sum_\ell N^3_\ell & = & 0, \\
\label{ttre}
{\rm Tr}([{\rm gravitational\,\,anomaly\,\,term}]^2[{\rm U}(1)_N]) & = & 0 : 
\nonumber \\
3(N_{Q_{1L}} + 2N_{Q_{\alpha L}}) - 
3(N_{U_{1,\alpha}} + N_{D_{1,\alpha}}) + \nonumber \\
N_{J_R} - 2N_{j_{\alpha R}} - N_{u^\prime_R}
-2N_{d^\prime_{\alpha R}} + \sum_\ell N_\ell & = & 0
\label{quttr}
\end{eqnarray}
\end{mathletters}
where $N_{U_{1,\alpha}}$ and $N_{D_{1,\alpha}}$ 
are the U(1)$_N$ quantum numbers of the standard 
quark flavors. The Witten global anomaly~\cite{Witten} does not involve the 
U(1)$_N$ quantum numbers. 
The three leptonic classical constraints are
\begin{equation}
N_\ell = 0
\label{ccqqe}
\end{equation}
coming from the Abelian gauge invariance of ${\cal L}_{\rm Y}$ where the 
electric charge quantization of the leptonic sector is already contained. 
From the quantum and classical gauge invariance constraints the electric 
charges of fundamental leptons and quarks, in the $|e|$ unit, 
are~\cite{DoffFP}
\begin{eqnarray}
{\cal Q}_{\nu_\ell} & = & 0, 
\quad {\cal Q}_\ell = -1, 
\quad {\cal Q}_D = -\frac{1}{3}, \nonumber \\
\quad {\cal Q}_U & = & -2{\cal Q}_D = -2{\cal Q}_{d^\prime_\alpha}
= {\cal Q}_{u_1}, \\
{\cal Q}_J & = & -5{\cal Q}_D, 
\quad {\cal Q}_{j_\alpha} = 4{\cal Q}_D \nonumber
\end{eqnarray}
even for massless neutrinos. There are no new constraints coming from 
the cancellation of mixed gauge and gravitational anomalies. 
If we consider the global symmetry associated to the conservation of 
the leptobaryon quantum number 
\begin{equation}
{\cal F} = L + B = \sum_\ell L_\ell + B\,,
\label{cclcio}
\end{equation}
which prevents neutrinos from getting a mass to be explicitly 
broken then Majorana mass terms arise if 
$\langle H^0_{1,4}\rangle\neq 0$ turning-off the VEVs of the 
$H^0_2$ and $H^0_3$ non-diagonal fields. However, turning-on 
the VEVs of all neutral scalar fields the neutrinos are 
Dirac--Majorana fermions~\cite{Espetal}. 
\par
Now let us add right-handed neutrinos as gauge flavor singlets. If 
$\langle H^0_1\rangle_0 = \langle H^0_2\rangle_0 = 
\langle H^0_4\rangle_0 = 0$, but $\langle H^0_3\rangle_0\neq 0$ for the 
charged lepton masses, 
the Dirac mass terms for the neutral fermions
\begin{equation}
-{\cal L}_{\rm Y}^{\nu_\ell,\langle\eta\rangle_0} = 
G_{f_\ell\nu_\ell}\bar f_{\ell L}\langle\eta\rangle_0\nu_{\ell^\prime R} 
+ {\rm H.c.}
\label{prprzz}
\end{equation}
arise in the Yukawa couplings through the multiplet of scalar fields
\begin{equation}
\eta = \left (
\begin{array}{c}
\eta^0_1 \\ 
\eta^-_1 \\ 
\eta^0_2 \\
\eta^+_2
\end{array}
\right )\sim ({\bf 1}_c,{\bf 4},0)
\label{sclf}
\end{equation}
also necessary in the quark sector, 
\begin{eqnarray}
-{\cal L}^{Q,\eta}_{\rm Y} & = & F_{1k}\bar Q_{1L}U_{kR}\,\eta \nonumber \\
& + & F^\prime_{\alpha k}\bar Q_{\alpha L} D_{kR}\,\bar\eta + {\rm H.c.}, 
\quad k=1,2,3
\label{cincque}
\end{eqnarray}
where
\begin{equation}
\bar\eta = \left (
\begin{array}{c}
\eta^+_2 \\ 
\eta^0_2 \\
\eta^-_1 \\
\eta^0_1 \\
\end{array}
\right )\sim ({\bf 1}_c,\bar{\bf 4},0).
\label{sseii}
\end{equation}
The classical gauge invariance of the 
${\cal L}_{\rm Y}^{\nu_\ell,\langle\eta\rangle_0}$ leptonic terms implies 
$N_{\nu_\ell} = 0$. The $L$ and $B$ attributions of the leptoquark 
fermions are $L_{j_\alpha} = -L_J = +2$, $B_J=B_{j_\alpha}=+\frac{1}{3}$, 
and the bilepton gauge bosons have lepton number $L=\pm 2$. The 
bileptons~\cite{CRa} are contained also in the stable-proton SU(15) grand 
unified theory~\cite{Lee90} with non-chiral fermions and anomaly 
cancellation through mirror fermions.
\par
The quantization of electric charge is inevitable in the $G_{3m1}$ 
models of leptoquark fermions and bilepton bosons~\cite{Dionetal,Cuyp} 
with three non-repetitive fermion generations breaking generation 
universality and does not depend on the character of the neutral fermions. 
Each generation is anomalous and are not replicas of one another 
so that the quantum anomalies cancel when the number of generations 
is a multiple of the number of color charges. 
There is not any connection between electric charge quantization and the 
massless Weyl, and the massive Dirac, Majorana or Dirac--Majorana 
fundamental neutral fermions. If the ${\cal F}$-symmetry is explicitly 
broken the neutrinos are Majorana fermions when 
$\langle H^0_{1,4}\rangle_0\neq 0$ and Dirac--Majorana fermions when 
all neutral components of the symmetric decuplet acquire their vacuum 
expectation values. This is another clue about the promising 
perspectives that neutrino address in the direction of new physics. 
The $G_{3m1}$ leptoquark-bilepton models have the 
leptonic representation content structure of grand unified theories in  
which the electric charge operator contains only the SU($m$)$_L$ diagonal 
generators whose number is the rank of the group. 
\acknowledgments
A.D. would like to thank the CAPES (Brazil) for full financial support.

\end{document}